\documentclass[prx, twocolumn, showpacs, preprintnumbers, amsmath, amssymb, superscriptaddress, aps,longbibliography]{revtex4-2}
\usepackage{amssymb}
\usepackage{latexsym}
\usepackage{graphicx}
\usepackage{hyperref}
\usepackage[english]{babel}
\hypersetup{linktocpage,,colorlinks=true}
\usepackage{color}
\usepackage{mathtools}
\usepackage{epsfig}
\usepackage{subfigure}
\usepackage{dcolumn}
\usepackage{amsmath}
\usepackage{amsfonts}
\usepackage{bm}

\newcommand{\dg}{^{\dagger}}
\newcommand{\be}{\begin{equation}}
\newcommand{\ee}{\end{equation}}
\newcommand{\bea}{\begin{eqnarray}}
\newcommand{\eea}{\end{eqnarray}}

\newcommand{\p}{\partial}
\newcommand{\s}{\sigma}

\newcommand{\la}{\langle}
\newcommand{\ra}{\rangle}
\newcommand{\rd}{\mbox{d}}
\newcommand{\ri}{\mbox{i}}
\newcommand{\re}{\mbox{e}}

\newcommand{\pmat}[1]{\begin{pmatrix}#1\end{pmatrix}}

\renewcommand{\vec}[1]{{\bm #1}}

\newcommand{\bk}{{\bf k}}
\newcommand{\bR}{{\bf R}}

\newcommand{\bx}{{\bf x}}
\newcommand{\by}{{\bf y}}

\begin{document}
\title{Perturbations in the  spin-orbital liquid (the Yao-Lee model)}
\author{ A. M. Tsvelik}
\affiliation{Division of Condensed Matter Physics and Materials Science, Brookhaven National Laboratory, Upton, NY 11973-5000, USA}
 \date{\today } 
 
 \begin{abstract} 
 I investigate the stability of the spin-orbital liquid described by the Yao-Lee model in the presence of multiple perturbations that naturally arise in a recently proposed microscopic realization. While most perturbations are irrelevant and do not qualitatively alter the exact solvability, we find that the Kitaev interaction plays a special role: it couples to the $  \mathbb{Z}_2  $ gauge field and renders the visons mobile. 
 
 \end{abstract}


\maketitle
Since P. W. Anderson's seminal proposal that doped spin liquids may naturally give rise to superconductivity, interest in quantum spin (and spin-orbital) liquids has remained strong. Beyond their potential for exotic superconductivity, these phases are fascinating in their own right due to their fractionalized excitations and emergent gauge fields.

Among the many theoretical models of spin liquids, exactly solvable ones occupy a privileged position as reliable reference points. The first and most prominent example is the Kitaev honeycomb model\cite{kitaev}. Generalizations soon followed, notably the Yao-Lee (YL) model\cite{YL} and its extensions proposed by Wu {\it et.al.} \cite{Wu2009} and the Dresden group.\cite{Vojta2} All these models can be formulated in terms of lattice Majorana fermions coupled to a static $  \mathbb{Z}_2  $ gauge field. They differ primarily in the number of Majorana species per site (one in the Kitaev model, three in the YL model, and an arbitrary number in the Dresden variant). This difference leads to distinct physical observables; for instance, spin operators are nonlocal in the Majorana representation of the Kitaev model but local in the YL model.

While these solvable models exhibit beautiful and rich physics—both in isolation and when coupled to conduction electrons in Kondo-lattice settings\cite{CT,CPT}—their experimental realization remains challenging. The Kitaev model, in particular, is highly sensitive to perturbations, which has fueled ongoing debate about the interpretation of experimental data in candidate materials.\cite{exp}
A potential realization of the Yao-Lee model faces a fundamental tension, as pointed out in Ref.~\cite{Kee2025}. Realizing the required exchange anisotropy typically demands strong spin-orbit coupling (SOC), while cleanly separating spin and orbital degrees of freedom requires weak SOC. To circumvent this, Ref.~\cite{Kee2025} proposed spatially separating the two: spin and orbital moments arising from $  e_g  $ orbitals (with weak SOC) reside on the honeycomb lattice sites, while the superexchange is mediated by intermediate anions possessing strong SOC. The resulting microscopic Hamiltonian contains the desired YL interaction together with additional couplings.
Subsequent work\cite{Onur} analyzed some of these extra interactions—specifically those that preserve the static nature of the $  \mathbb{Z}_2  $ gauge field. In this Brief Report, I employ analytic methods to examine the stability of the exact YL solution against all perturbations present in the microscopic model of Ref.~\cite{Kee2025}. I show that most are irrelevant, but the Kitaev term stands out as it couples to the gauge sector and makes the visons mobile.



 {\it The YL model.}
 
\begin{figure}[t]
\includegraphics[width=0.3 \textwidth]{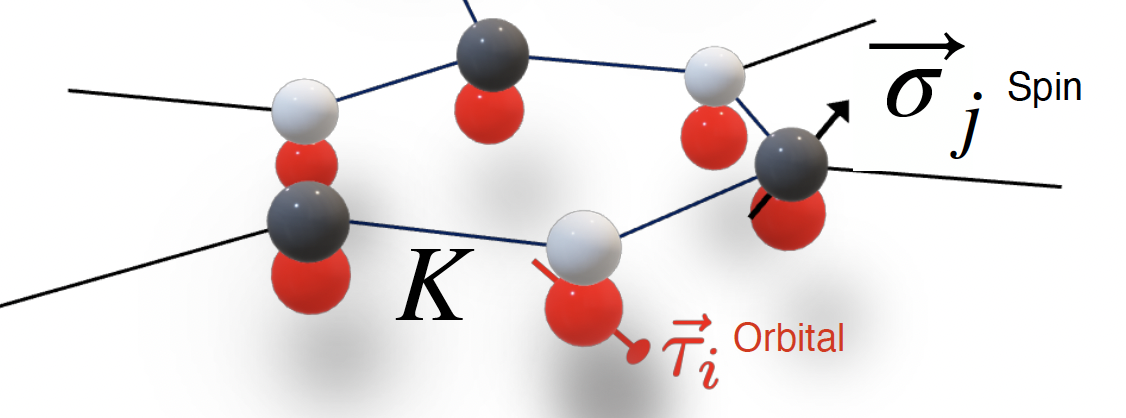}
\vspace{-0.2cm}
\caption{Yao-Lee model on hexagonal lattice. }
\label{Fig:YL}
\vspace{-0.5cm}
\end{figure}

  The perturbed YL model is 
 \bea
 H = H_{YL} + \delta H, \label{model} 
 \eea
 where $\delta H$ stands for a perturbation (to be specified later) and $H_{YL}$ is the Hamiltonian of the Yao-Lee (YL) spin liquid \cite{YL}:
 \bea
 H_{YL} = - (K/4)\sum_{\la i,j\ra}(\vec\s_i\vec\s_j)\tau_i^{\alpha_{ij}}\tau_j^{\alpha_{ij}}, \label{ksl}
 \eea
 where indices $i,j$ denote a pair of neighboring sites. The model is solvable on any lattice with coordination number three. In this note I  will confine myself to  hexagonal lattice. 
 $\s$ and $\tau$ are Pauli matrices acting in the spin and the orbital space respectively. Indices $\alpha$ mark $x,y,z$ bonds of the hexagonal lattice (see Fig. \ref{Fig:YL}). With $\s$ fixed the interaction of $\tau$'s would be exactly like in the Kitaev model. 
 As the Kitaev one, the Yao-Lee  model admits a representation in terms of noninteracting Majorana fermions moving in a background of static Z$_2$ gauge field  $u_{ij}$ \cite{YL}:
 \bea
 H_{YL} = \ri K\sum_{\la i,j\ra} u_{ij}(\vec\chi_i \vec\chi_j). \label{majoranas}
 \eea 
 Excitations of this field are static fluxes (visons), their creation requires energy; the vison gap for the YL model is three times greater than the gap for the Kitaev model \cite{kitaev,Vgap2}: $\Delta_v \approx 0.39 K$.
 
 To see this we need to use the  fermionization formulae for the Pauli matrices:
 \bea
\vec{\s}_{j} =-\ri \vec{\chi}_{j}\times
\vec{\chi}_{j}, ~~\vec{\tau }_{j}=-\ri \vec{b}_{j}\times
\vec{b}_{j}, \label{fermionization}
 \eea

 These satisfy canonical anti-commutation algebras
$\{\chi^{a}_{j},\chi^{b}_{j} \}=\{b^{a}_{j},b^{b}_{j} \}
= \delta^{ab}\delta_{ij}$ and $\{\chi^{a}_{j},b^{b}_{k} \}= 0$.
There is a constraint 
\begin{equation}\label{constraint}
\hat D_{j}=8i
\chi_{j}^{1}\chi_{j}^{2}\chi_{j}^{3}b_{j}^{1}b_{j}^{2}b_{j}^{3}
\end{equation}
Now $\hat D_{j}$, with eigenvalues $D_{j}=\pm 1$,
commutes with $H$, and the constraint
$D_{j}=1$ 
selects a set of physical states 
\begin{equation}\label{constraint2}
\vert \psi_{p}\rangle = \prod_{j}\frac{1}{2} (1+D_{j})\vert \psi \rangle .
\end{equation}
In this physical subspace 
\begin{equation}\label{stau}
\s_{j}^{a}\tau_{j}^{\alpha }= -2i \chi_{j}^{a}b^{\alpha }_{j}.
\end{equation}
which enables us to rewrite \eqref{ksl} as \eqref{majoranas} 
where 
$ \hat  u_{ij}=2i 
b_{i}^{\alpha_{ij}}b_{j}^{\alpha_{ij}}$ are bond variables 
with eigenvalues $u_{ij}=\pm 1$. The $\hat u_{ij}$
 commute with the full Hamiltonian $H$, 
forming static $Z_{2}$ gauge
fields.

{\it YL model perturbed by orbital interaction. Mean field approach.}  Below I will consider analytically all perturbations which were treated numerically in Ref. \cite{Kee2025}. I start with those which where also considered analytically in \cite{Onur}: 
  \bea
  && \delta H_H = (J_H/4) \sum_{<i,j>}(\vec\s_i\vec\s_j), \label{Heis}\\
  && \delta H_K = (\tilde K/8)\sum_{<i,j>}\tau_i^{\alpha_{ij}}\tau_j^{\alpha_{ij}}.\label{Kit}
  \eea
  Both of them can be expressed in terms of the Majoranas and the static gauge field $u_{ij}$. Using (\ref{fermionization}) and constraint (\ref{constraint}) we get
  \bea
  && \delta H_H = J_H \sum_{<i,j>}\sum_{a>b} (\chi^a_i\chi^a_j)(\chi^b_i\chi_j^b), \\
  && \delta H_K = \ri\tilde K \sum_{<i,j>}u_{ij}(\chi^1\chi^2\chi^3)_i(\chi^1\chi^2\chi^3)_j \label{tildeK}
\eea
These are irrelevant perturbations which do not violate stability of the YL ground state. One comment concerning $\delta H_K$ is in order. According to \cite{Onur} with increase of $\tilde K$ the system undergoes the T=0 phase transition to a state with a partial magnetic order. In \cite{Onur} this transition was treated by mean field. The triad of Majorana fermions spontaneously separated into the group of $1+2$ with two fermions forming the magnetization and the remaining one describing the Kitaev sector. The  limit of $K=0$ (the pure Kitaev model) is particularly illuminating since in this limit any pair of local Majoranas commutes with the Hamiltonian and can be identified with $\s^z$(for instance, one take $\s^z = 2\ri\chi^2\chi^3$). Then the remaining Majorana (in the given case it would be $\chi^1$) is identified with the Kitaev fermion $\chi^0$. However, at finite $K$ $\s^z$ no longer commutes with the Hamiltonian though its partial fermionization is still possible. Namely, fermionizing the $\tau$-operators we  rewrite the entire Hamiltonian (\ref{ksl},\ref{tildeK}) as 
\bea
&&H_{YL} + \delta H_K = \nonumber\\
&& \ri \sum_{<i,j>}u_{ij}\Big[(\tilde K/2) + K(\vec\s_i\vec\s_j)\Big]\chi^0_i\chi^0_j \label{largetildeK}
\eea
Then since $\s^z$ no longer commutes with the Hamiltonian,  $\chi^0$ should be identified with a three-body  bound state of the YL Maojoranas. The $\tilde K$ term in (\ref{largetildeK}) fixes the ground state average of $u_{ij}\chi^0_i\chi_j^0$ generating  the Heisenberg exchange for the $\s$'s. This description works not only for the ordered state that exists just at $T=0$, but at finite temperatures $T<< \tilde K$. 

\begin{figure}[t]
\includegraphics[width=0.2 \textwidth]{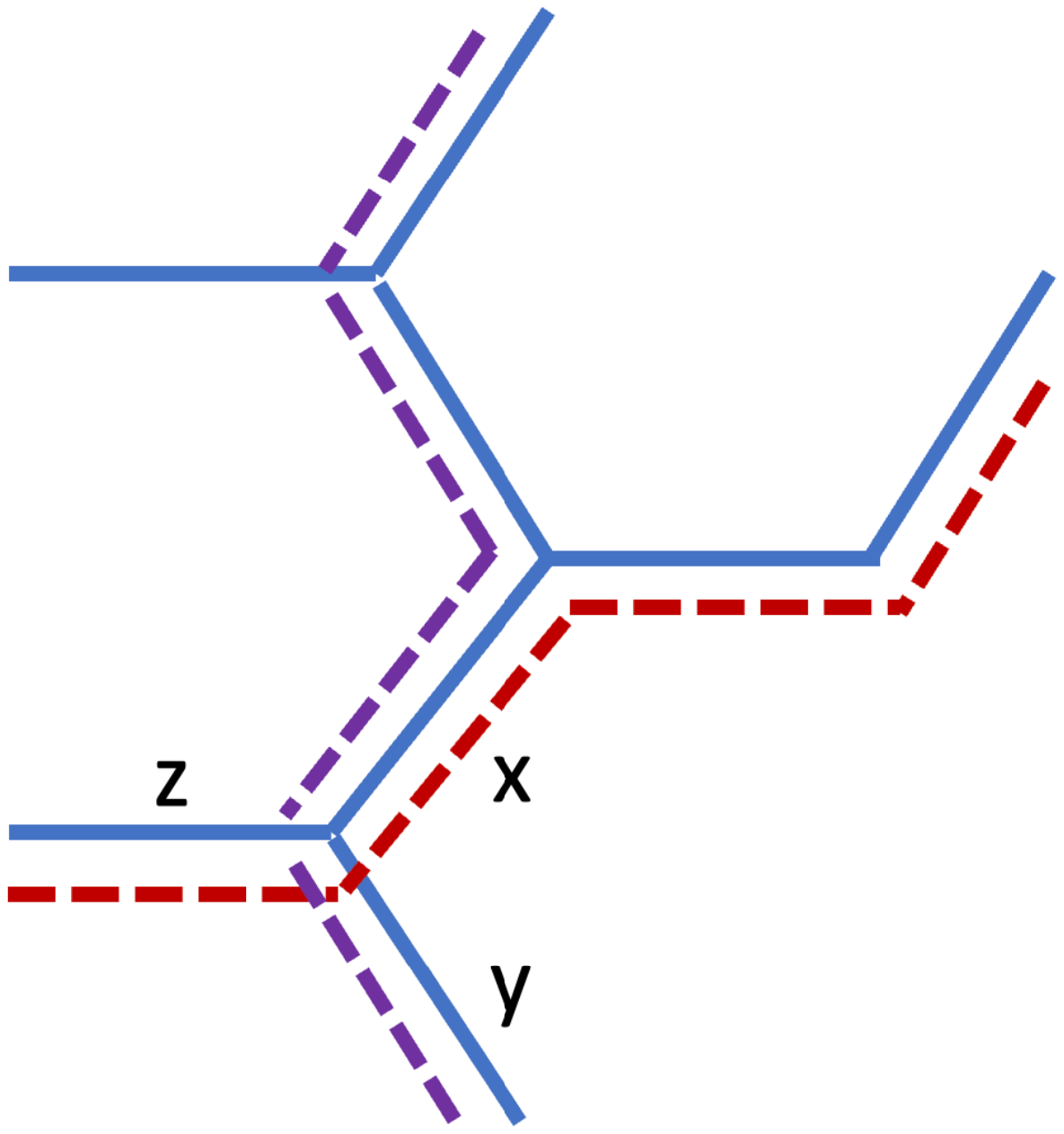}
\vspace{-0.2cm}
\caption{Paths of propagation of the $b^{\alpha}$-fermions  generated by perturbation (\ref{pert}). The $y$-fermion follows the red and the $z$-fermion follows the purple path.}
\label{Fig:Path}
\vspace{-0.5cm}
\end{figure}

  As will be immediately clear  the most dangerous perturbation is \bea
  \delta H = (J/4)\sum_{<i,j>} (\vec\tau_i\vec\tau_j). \label{pert}
 \eea
 This perturbation makes the gauge field dynamic which has not been demonstrated in the previous works on the YL model. 
 Using the fermionization rules (\ref{fermionization}) we rewrite it as 
 \bea
 \delta H = J\sum_{<i,j>}\sum_{\alpha > \beta}(b_i^{\alpha}
 b_j^{\alpha})(b^{\beta}_i b_j^{\beta}). \label{deltatau}
 \eea
 In what follows we will denote the links of the lattice as $x,y,z$.  It turns out  that this perturbation contains a part bilinear in the $b$-fermions. This is due to the fact that in the original YL model some of the $b-b$ pairs condense as the $Z_2$ gauge field. Let us take, for instance, link $x$ and  consider a perturbation around the vacuum where $u_{i, i+e_x} = 2\ri\la b^x(i)b^x(i+e_x)\ra$. Then in the first approximation the perturbation (\ref{deltatau}) on this link is 

 \bea
 &&\delta H(i,i+e_x)  \approx \\
 && \frac{\ri }{2} J u_{i,i+e_x}[b^y(i)b^y(i+e_x) + b^z(i)b^z(i+e_x)] +..., 
 \eea
where dots stand for the quartic terms. From this we may conclude that (\ref{deltatau}) generates such hopping matrix elements  that the $y$-fermion will propagate avoiding $y$-links and the $z$-fermion will avoid $z$-links as shown on Fig. {\ref{Fig:Path}}. This process liberates the $b$-fermions which were confined each on its own link. As far as the hopping matrix element on the own link is concerned, we can estimate it as follows. When we flip the sign the gauge field on a link, we create two visons. Therefore  we should assign to the fermions confined to a particular link a hopping matrix element equal to twice the vison energy $2\Delta_v$. The resulting effective Hamiltonian for $b$-fermions is then the anisotropic Kitaev model:
\bea
\delta H_{eff} = \frac{\ri}{2} \sum_{<i,j>} J^a b^a(i)u_{ij}b^a(j), \label{Eff}
\eea
where for each flavor $a$ two links have matrix element $J$ and one $2\Delta_v$, where $\Delta_v$ is the vison energy.
 Diagonalizing (\ref{Eff}) we obtain the dispersion  of the $y$-fermion:
 \bea
 E_y(k) = \Big|2\Delta_v + \ri J\re^{-3\ri {\bf e}_y{\bf k}/2}\sin{|{\bf k}\times{\bf e}_y|}\Big|,
  \eea
   Likewise we get dispersions of the other two $b$-Majoranas by replacing $y$ with $x,z$. The gap vanishes at $2\Delta_v = J$, this gives estimate for the stability of the spin liquid ground state. At $2\Delta_v > J$ the minima of the dispersion are located at $\pm (\pi/\sqrt 3, \pi/3)$ and the points rotated by $\pm 2\pi/3$ angle. 

Our results are in full agreement with previous works\cite{Kee2025,Onur} in that the Yao-Lee (YL) model remains stable against small perturbations corresponding to the Heisenberg interaction between the $  \mathbf{s}  $-spins [Eq.~(\ref{Heis})] and the Kitaev interaction between the $  \boldsymbol{\tau}  $-spins [Eq.~(\ref{Kit})].
However, in contrast to the numerical findings of Ref.~\cite{Kee2025} reflected on Fig. 3 in this paper, I identify a finite window of stability against the Heisenberg interaction between the $  \boldsymbol{\tau}  $-spins—the degrees of freedom that generate the $  \mathbb{Z}_2  $ gauge structure in the YL model. A possible reason for disagreement is a relative smallness of the stability window. 
This particular perturbation liberates the $  b  $-Majorana fermions, which in the pure YL model are confined to the lattice links and form the static gauge field. For sufficiently small perturbation strength (compared to the vison gap), the $  b  $-Majoranas become deconfined while the YL ground state itself remains stable. Because the $  b  $-fermions acquire a finite mass, the resulting phase is a gapped topological spin-orbital liquid. A conceptually similar situation, in which perturbations render the visons mobile, was previously analyzed for the Kitaev model in Ref.~\cite{Rosch}.

   {\it Acknowledgements}

 I am grateful to Onur Erten for a productive discussion. This work was supported by Office of Basic Energy Sciences, Material
 Sciences and Engineering Division, U.S. Department of Energy (DOE)
 under Contracts No. DE-SC0012704.  

 \bibliographystyle{apsrev4-2}
\bibliography{flstar.bib}

\end{document}